 \documentclass[preprint,aps,showpacs,nofootinbib]{revtex4}
\usepackage{graphicx,amsmath,amssymb}
\usepackage{psfrag}
\usepackage{color}
\usepackage{xcolor}
\usepackage{hyperref}
\usepackage{slashed}
\usepackage{ulem}
\usepackage{cancel}
\newcommand{\be}{\begin {eqnarray}}
\newcommand{\ee}{\end{eqnarray}}

\usepackage[T1]{fontenc}
\usepackage[utf8]{inputenc}
\usepackage{lmodern}
\usepackage{amsfonts}
\usepackage{color}
\newcommand{\CB}{Borici-Creutz}
\newcommand{\p}{\pi_{\Gamma}}
\newcommand{\s}{\sigma}

\newcommand\redout{\bgroup\markoverwith{\textcolor{blue}{\rule[0.5ex]{6pt}{1.0pt}}}\ULon}
\begin{document}
\title{  Mass spectroscopy using Borici-Creutz fermion on  2D lattice}
\author{ J. Goswami$^1$,  D. Chakrabarti$^1$,  S. Basak$^2$ }
\affiliation{  $^1$Department of Physics,  Indian Institute of Technology Kanpur, Kanpur - 208016, India\\
$^2$ School of Physical Sciences, NISER, Bhubaneswar - 751005, India.}
\date{\today}
\begin{abstract}
Minimally doubled fermion proposed by Creutz and Borici is a promising chiral fermion formulation on  lattice. In this work, we present  excited state mass spectroscopy for the meson bound states in Gross-Neveu model using  Borici-Creutz fermion. We also evaluate the effective   fermion mass as a function of coupling constant  which shows a chiral phase transition at strong coupling. The lowest lying meson in 2-dimensional QED is also  obtained using  Borici-Creutz  fermion. 
 \end{abstract}

\pacs{11.15.Ha, 11.10.Kk,  11.30.Rd,12.40Yx}
\maketitle
\section{Introduction}
Chiral fermion formulation is always a challenging task on lattice and minimally doubled fermions in recent times have drawn attention as promising lattice formulations of chiral fermion. Karsten\cite{K} and Wilczek\cite{W} proposed one formulation of minimally doubled fermion and another was developed by Creutz\cite{creutz} and Borici\cite{borici}.  Both the formulations break the hypercubic symmetry on the lattice \cite{bedaque} and thus allow non-covariant counter terms. The important  question is how bad the effects of the symmetry breaking are in a numerical simulation. It was  shown that a consistent renormalizable theory for minimally doubled fermion can be constructed by fixing only three counter terms allowed by the symmetry and the counter terms for BC action at  one loop in perturbation theory have been evaluated \cite{capitani1,capitani2}.
But, till date, sufficient numerical studies of the minimally doubled fermions have not been done. The purpose of this work is to investigate  Borici-Creutz(BC) formulation numerically  in some models. \\
The BC fermion formulation was motivated by the fact that electrons on  graphene lattice are described by a massless quasi-relativistic Dirac equation. The  BC fermion describes two chiral modes or the two flavors of chiral fermion  located at $(0,0,0,0)$ and at $(\frac{\pi}{2},\frac{\pi}{2},\frac{\pi}{2},\frac{\pi}{2})$.
It was shown that in presence of gauge background with integer-valued topological charge, BC action satisfies the Atiyah-Singer index theorem\cite{dc}. In \cite{GCB}, using BC fermion  we have shown a chiral phase transition in the Gross-Neveu model.  We extend that investigation further in this work to  meson spectroscopy of the Gross-Neveu model  as well as in 2D QED (QED$_2$) using BC fermion. Chiral and parity-broken(Aoki) phase structures of the Gross-Neveu model have been studied for Wilson  and Karsten-Wilczek fermions \cite{CKM,misumi}. A lattice simulation of the Gross-Neveu model   using  the Wilson fermion was done by Korzec et al\cite{korzec}, where the recovery of chirally invariant Gross-Neveu model from a lattice model was studied.
The semimetal-insulator phase transition on a graphene lattice with Thirring type four fermion interactions has been studied by Hands and collaborators\cite{hands} and the strong coupling analysis of  the tight-binding graphene model with Kekule distortion term has been done by Araki\cite{araki}. We use pseudofermion HMC for lattice simulation of the Gross-Neveu model. The feasibility of psedofermion algorithm in the Gross-Neveu model has been studied by Campostrini {\it et. al.}\cite{Camp}.

In this paper, we perform  hybrid Monte Carlo(HMC) simulation to  investigate the excited state  spectrum of the lattice Gross-Neveu model.  Extraction of the excited state spectrum is always a difficult task on lattice. As of now, variational method gives the best spectrum. From the slope of the correlators, we first obtain  some preliminary estimate about the masses and then we use variational method to extract the meson masses. Excited state spectroscopy in the Gross-Neveu model with Wilson fermion has been studied in \cite{Danzer}, where the authors
obtained the ground state as the only bound state and the other excited states were scattering states. With the BC fermion, we have obtained three states,  two of them are bound states (ground state and one excited state) and the third one appears to be a scattering state. We also evaluate the fermion mass in the model which shows that it is consistent with a chiral phase transition at large coupling   observed in \cite{GCB}.
Next we investigate the meson mass spectrum in QED in two dimension.  QED$_2$ having confinement has bound state spectrum  and  serves as a toy model for hadronic bound states in QCD.  In continuum, massless Schwinger model is exactly solvable and can be represtented as a free boson theory.  QED$_2$ or Schwinger model has been  studied to great extent on lattice (see \cite{Gutsfeld,Gattringer} and references therein).  Schwinger model using Hamiltonian formalism on lattice has been investigated in \cite{cichy}.  QED$_2$ also serves as a good toy model for numerical study of chiral  fermions. A 2-flavor
Schwinger model with light fermions have been studied with    dynamical  overlap fermion \cite{ Giusti, Hip}.  
Here, we study the  model with the minimally doubled fermion, namely, the BC fermion.



\section{Spectroscopy of the Gross-Neveu model}
The free BC action in 2D  is written as,
  \be
  S&=&\sum_{n}\left[\frac{1}{2}\sum_{\mu}\overline{\psi}_n\gamma_{\mu}(\psi_{n+\mu}-\psi_{n-\mu})-\frac{i r}{2}\sum_{\mu}\overline{\psi}_n(\Gamma-\gamma_{\mu})(2\psi_{n}-\psi_{n+\mu}-\psi_{n-\mu}) \right.\nonumber \\
 && \left. -i(2-c_{3})\overline{\psi}_{n}\Gamma\psi_{n}+m_0\overline{\psi}_{n}\psi_{n}\right]\label{CB},
  \ee
  where, $\mu=1,2$ and  $\Gamma=\frac{1}{2}(\gamma_{1}+\gamma_{2})$ satisfies  $\{\Gamma,\gamma_{\mu}\} =1$.
Including four-fermion interactions, the  Gross-Neveu model with a discrete $\gamma_5$ symmetry on  lattice is given by
     \begin{eqnarray}
       S&=&\sum_{n}\left[\frac{1}{2}\sum_{\mu}\overline{\psi}_n\gamma_{\mu}(\psi_{n+\mu}-\psi_{n-\mu})-\frac{i r}{2}\sum_{\mu}\overline{\psi}_n(\Gamma-\gamma_{\mu})(2\psi_{n}-\psi_{n+\mu}-\psi_{n-\mu}) \right.\nonumber \\
        && \left. -i(2-c_{3})\overline{\psi}_{n}\Gamma\psi_{n}+m_0\overline{\psi}_{n}\psi_{n}-\frac{g^{2}}{2N}[(\overline{\psi}_{n}\psi_{n})^{2}+(\overline\psi_{n}i\Gamma\psi_{n})^{2}\right],\label{gnmodel}
        \ee
where,  $g$ is the coupling constant which we consider  the same for both  scalar and vector four point interactions and we set $r=1$ in our calculations. Since the parity is broken by the BC action, a counter term $c_3$ is added to it. Detailed discussion about the  $c_3$-term can be found in \cite{GCB}.
Now,  the action is rewritten explicitly in terms of the auxiliary fields as
   \begin{equation}
     S=\sum_{m,n}\overline{\psi}_{m}M_{mn}\psi_{n}+{\frac{N}{2g^{2}}}(\sigma^{2}+\pi_{\Gamma}^{2}),
   \end{equation}
   where $N$ is the number of flavors. The auxiliary fields 
   \begin{eqnarray}
  \s=-\frac{g^2}{N}(\overline{\psi}\psi), \nonumber\\
  \p=-\frac{g^2}{N}(\overline{\psi}i\Gamma\psi)
\end{eqnarray}
are defined  in the dual lattice sites $\tilde{l}$ surrounding the direct lattice site $l$ \cite{hands2}.
 \be 
 M_{mn}=D_{mn}+\frac{1}{4}\sum_{\langle l,\tilde{l}\rangle}{(\sigma(\tilde{l})+i\pi_{\Gamma}(\tilde{l})\Gamma)},
 \ee  
  where $\langle l, \tilde{l}\rangle$ denotes  four dual lattice sites $\tilde{l}$ surrounding the direct lattice site $l$ and 
  $D_{mn}$ is the BC Dirac operator:
 \be 
 D_{mn}&=&\frac{1}{2}\gamma_\mu(\delta_{n,m+\mu}-\delta_{n,m-\mu})+\frac{i}{2}(\Gamma-\gamma_\mu)(\delta_{n,m+\mu}+\delta_{n,m-\mu})\nonumber\\
 &&~~~ -((2-c_3)i\Gamma-m_0)\delta_{m,n}.\label{bcdirac}
 \ee
Since $M$ is a complex matrix, we work with (${M}^{\dagger}M$) to make it real and positive definite and  integrate out the fermion fields by the pseudofermion method.
 Since  the {\CB} action describes two flavors, the number of flavors becomes double i.e., $N_f=2N=4$ for an action with  ($M^\dagger M$).  In 2D, counter terms can be tuned to restore Lorentz invariant dispersion relation and here we can set the dimension two counterterm to zero; Lorentz symmetry can be controlled by $c_3$ alone. The minimally doubled fermions have physical dispersion relation  for $0<c_3<0.59$ and $3.41<c_3<4$ and the two chiral phase boundaries are near $c_3=0$ and $c_3=4$ \cite{GCB}. For the mass spectroscopy, we consider here $c_3=0.1$ i.e.,   in the  minimal double region with Lorentz invariant dispersion relation and  reasonably away from the phase boundaries. The value of $c_3$ is not completely arbitrary, other values of $c_3$ in the minimally double region are equally valid, the numerical values may change  for different $c_3$, but the physics remains unchanged. 
 
 With pseudofermions the action becomes ,
 \begin{equation}
  S=\phi^{\dagger}(M^{\dagger}M)^{-1}\phi+\beta(\sigma^{2}+\pi_{\Gamma}^{2}),
\end{equation}
where, $\beta={1/ {g^{2}}}$.
  We perform hybrid Monte Carlo(HMC)  simulation using this lattice action. 
 The  configurations are generated  by considering step-size($\triangle {t}$)=0.1 in the leapfrog method and ten
 steps per trajectory in the molecular dynamics chain. We do not use any preconditioning during the simulation.
  First 1000 ensembles  are rejected for thermalization and analysis is performed over the next 8000 ensembles.

\subsection{Correlators }
For meson mass spectrum calculation, we need to evaluate the correlators
\be
C_{ij}(t)=\langle O_i(t)O_j^\dagger(0)\rangle.
\ee
Since  we cannot have orbital angular momentum in 2D, the interpolators ($O_i$) are labelled by  parity  and charge conjugation only.  It is important to choose the appropriate operators which have good overlaps with the low lying states.  For the meson spectroscopy,  we consider only the odd parity interpolators. The even parity interpolators are not considered as they  do not decay exponentially and correspond to condensates\cite{Danzer}.  Since under parity $\psi(x,t)\to \gamma_2 \psi(-x,t)$, the odd parity  interpolators can be constructed with $\gamma_1$ or $\gamma_5$. Along with the  local source, one can also construct the interpolators with the fields at  different lattice sites shifted along the spatial direction ie., with $\psi(x\pm n,t)$ . If one considers a relative negative sign in between $\psi(x+n,t)$ and $\psi(x-n,t)$ then this corresponds to a derivative source which are found to be important for excited state spectroscopy\cite{Danzer,Dsource}.  Combining the field operators at different lattice sites, many interpolators can be constructed but it was found in our numerical analysis that  they mostly couple to the ground state. In\cite{Danzer},  a set of nine different interpolators were listed.
 Here we  list some of the parity odd interpolators for the GN model which we expect to couple to ground state as well as excited states:
\be
O_1(t)&=&{\overline \psi}(x,t)\gamma_5\psi(x,t) \nonumber\\
O_2(t)&=&\frac{1}{4}\big(({\overline \psi}(x+m,t)-\overline{\psi}(x-m,t)\big)\gamma_5\big(\psi(x+n,t)-\psi(x-n,t)\big), (m=3, n=3) \nonumber\\
O_3(t)&=&\frac{1}{4}\big(({\overline \psi}(x+m,t)-\overline{\psi}(x-m,t)\big)\gamma_5\big(\psi(x+n,t)-\psi(x-n,t)\big), (m=5, n=3)\nonumber\\
O_4(t)&=&\frac{1}{4}\big(({\overline \psi}(x+m,t)-\overline{\psi}(x-m,t)\big)\gamma_1\big(\psi(x+n,t)-\psi(x-n,t)\big), (m=4, n=3)\nonumber\\
O_5(t)&=&\frac{1}{4}\big(({\overline \psi}(x+m,t)+\overline{\psi}(x-m,t)\big)\gamma_1\big(\psi(x+n,t)-\psi(x-n,t)\big), (m=5, n=3),\nonumber\\
\label{interpol}
\ee
where sum over $x$ is implied in order  to have  zero momentum  states and $\gamma_5=i\gamma_1\gamma_2$. All the interpolators are odd under $C$-parity ($C=-1$). With different values of $m$ and $n$, we can have different interpolators, but the one that are found to couple with ground state as well as the excited states are for the values listed above in Eq.(\ref{interpol}), other interpolators do not couple to new states but only reproduce the similar results.

\begin{figure}[htbp]
\begin{center}
\small{(a)}\includegraphics[width=8cm,clip]{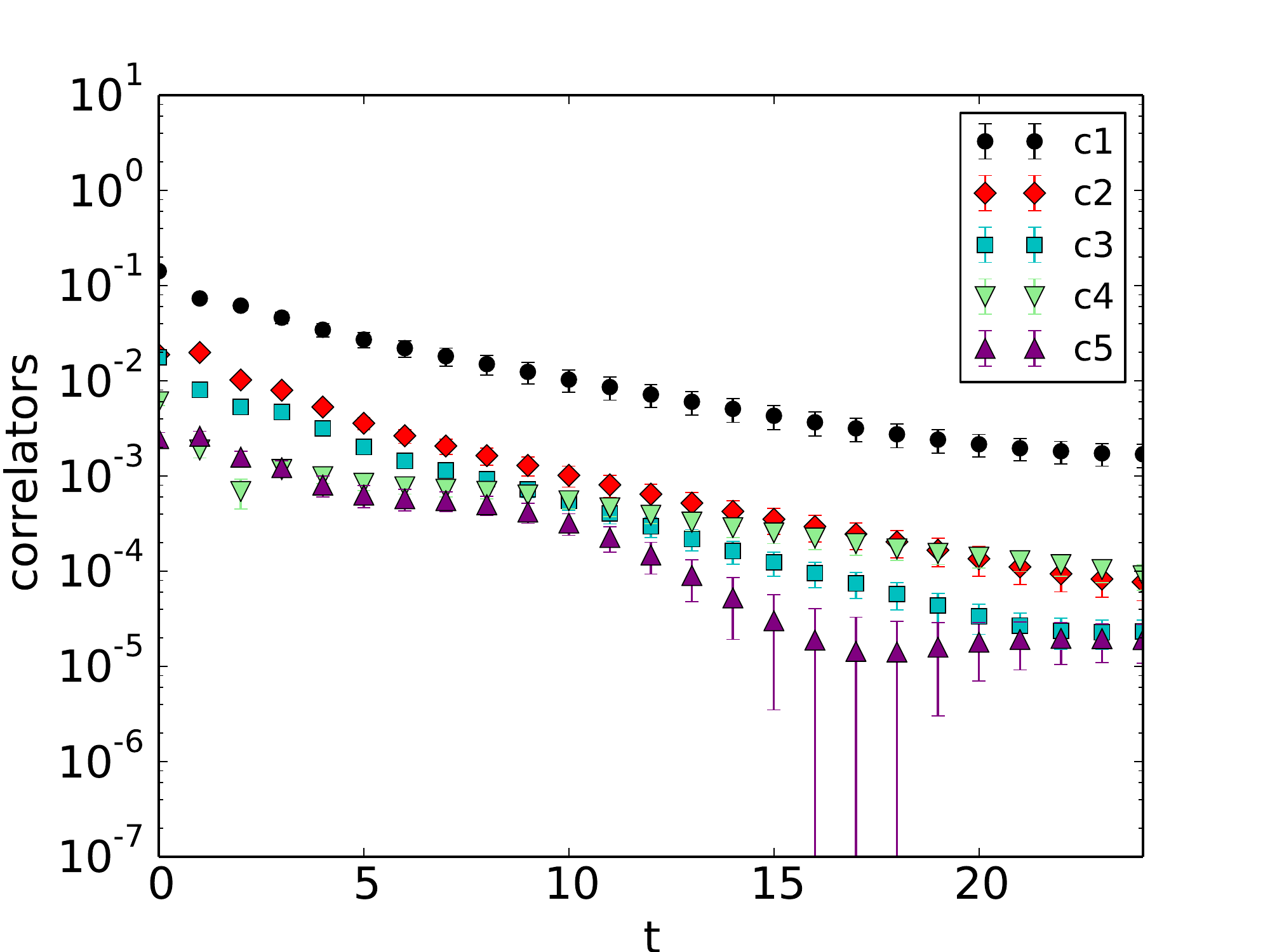}
\small{(b)}\includegraphics[width=8cm,clip]{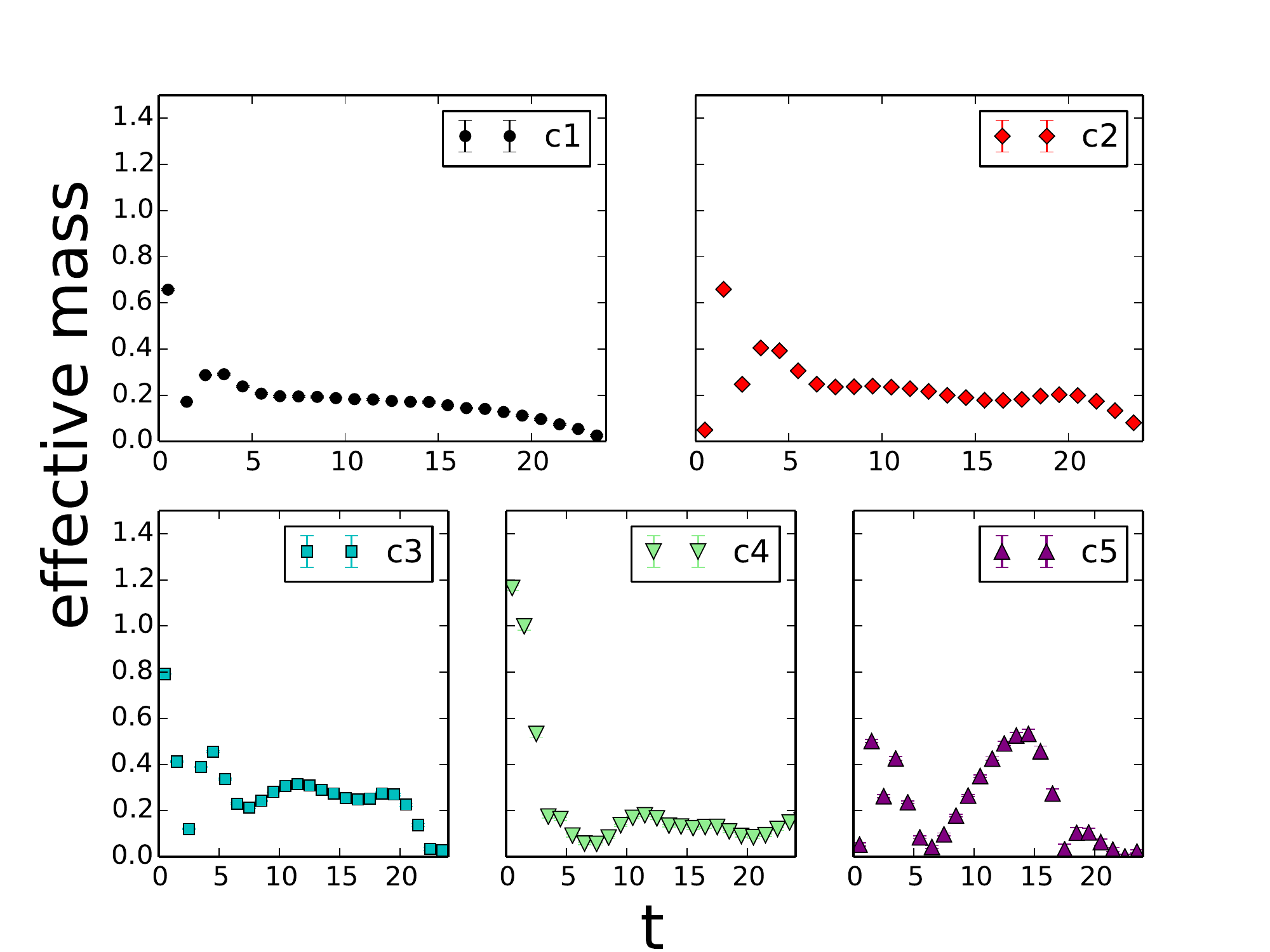}
\caption{ Diagonal correlators (in the plots $c1\equiv C_{11}$, etc.) 
and effective mass of meson in GN model for $16\times 48$ lattice\label{GN_meson_corr}}
\end{center}
\end{figure}

\begin{figure}[htbp]
\begin{center}
\small{(a)}\includegraphics[width=8cm,clip]{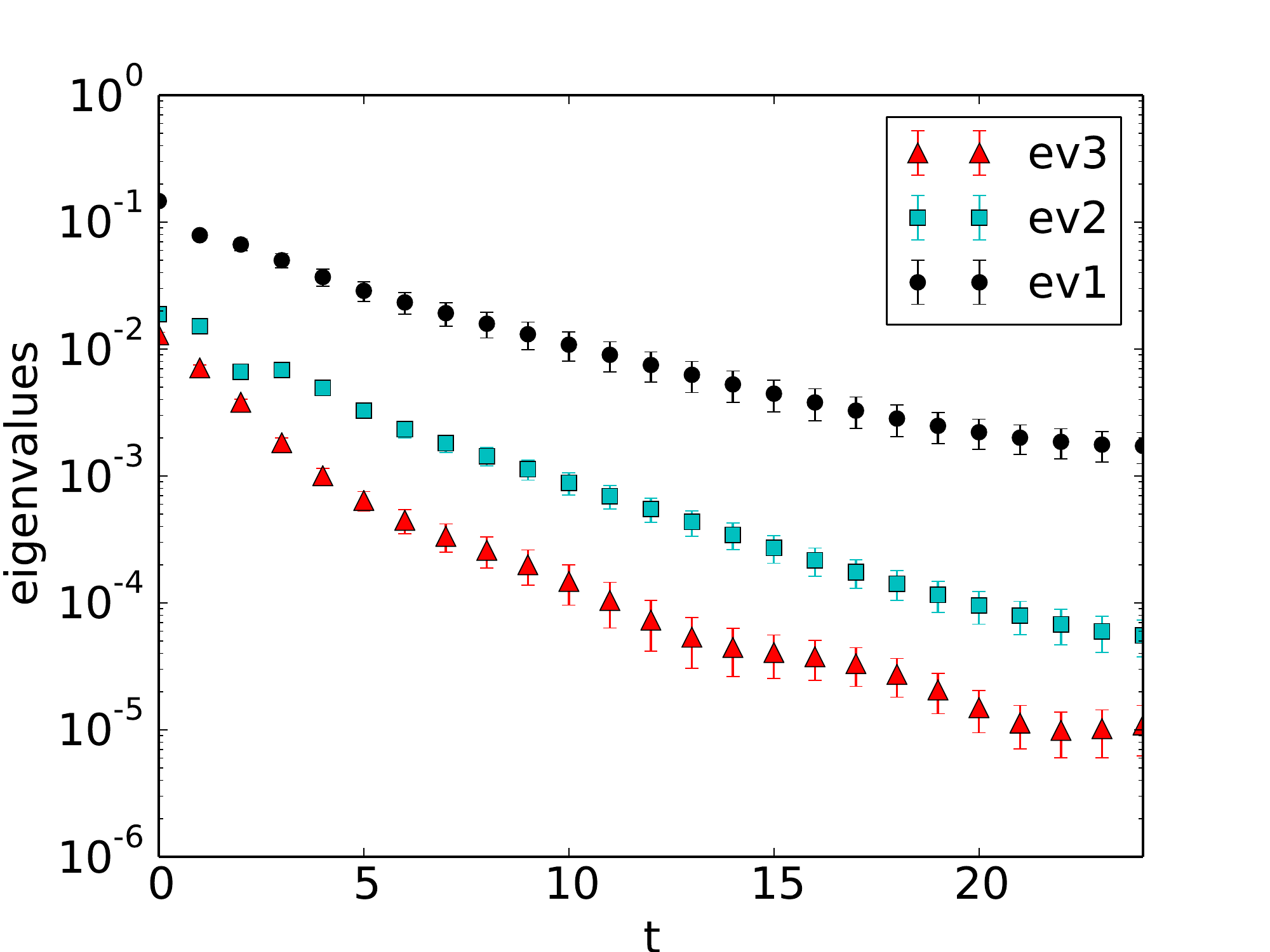}
\small{(b)}\includegraphics[width=8cm, clip]{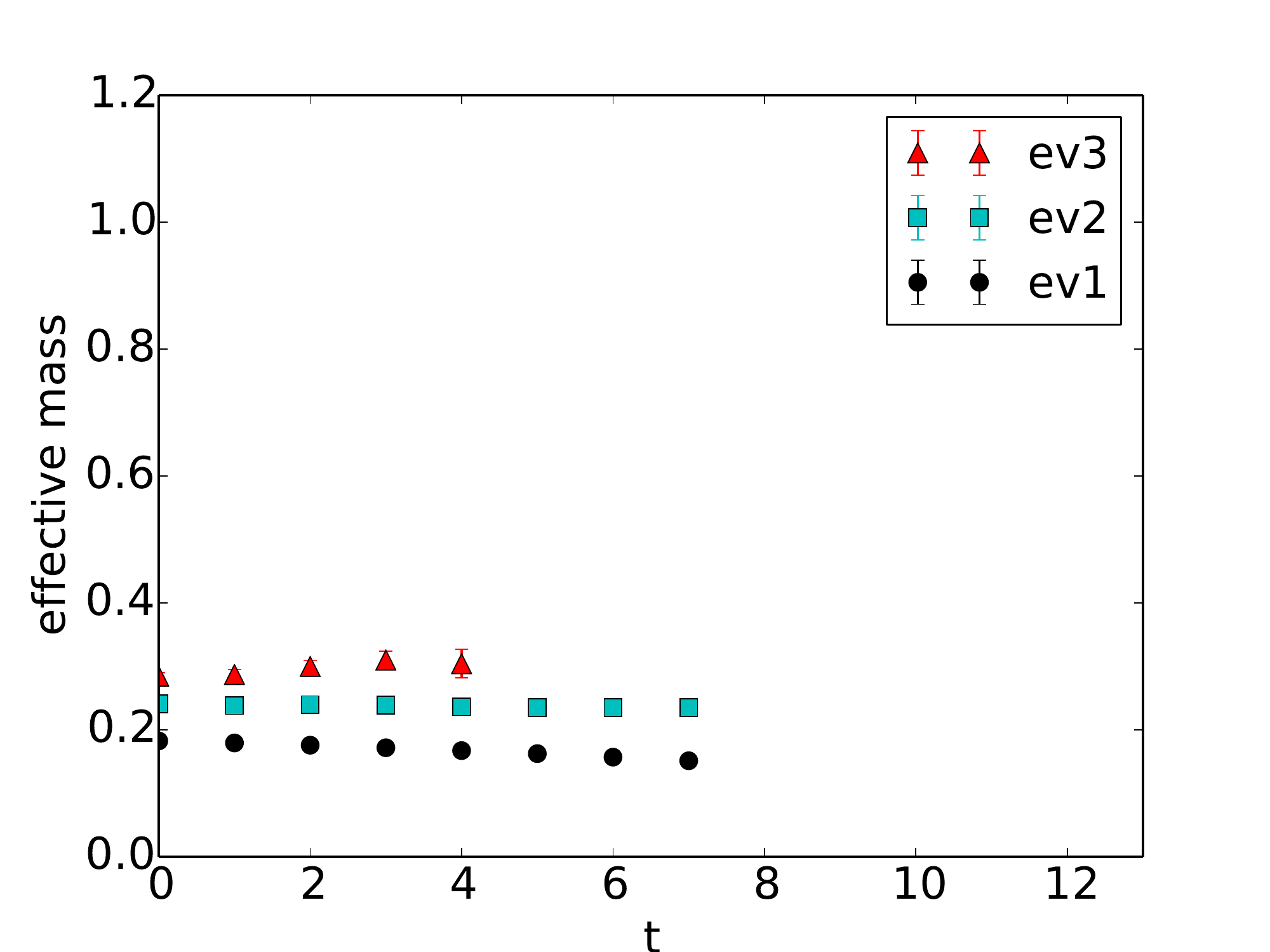}
\caption{ Eigenvalues and effective mass of the correlators for $16 \times 48$ lattice  \label{GN_eigen_meff}}
\end{center}
\end{figure}

\begin{figure}[htbp]
\begin{center}
\small{(a)}\includegraphics[width=8cm,clip]{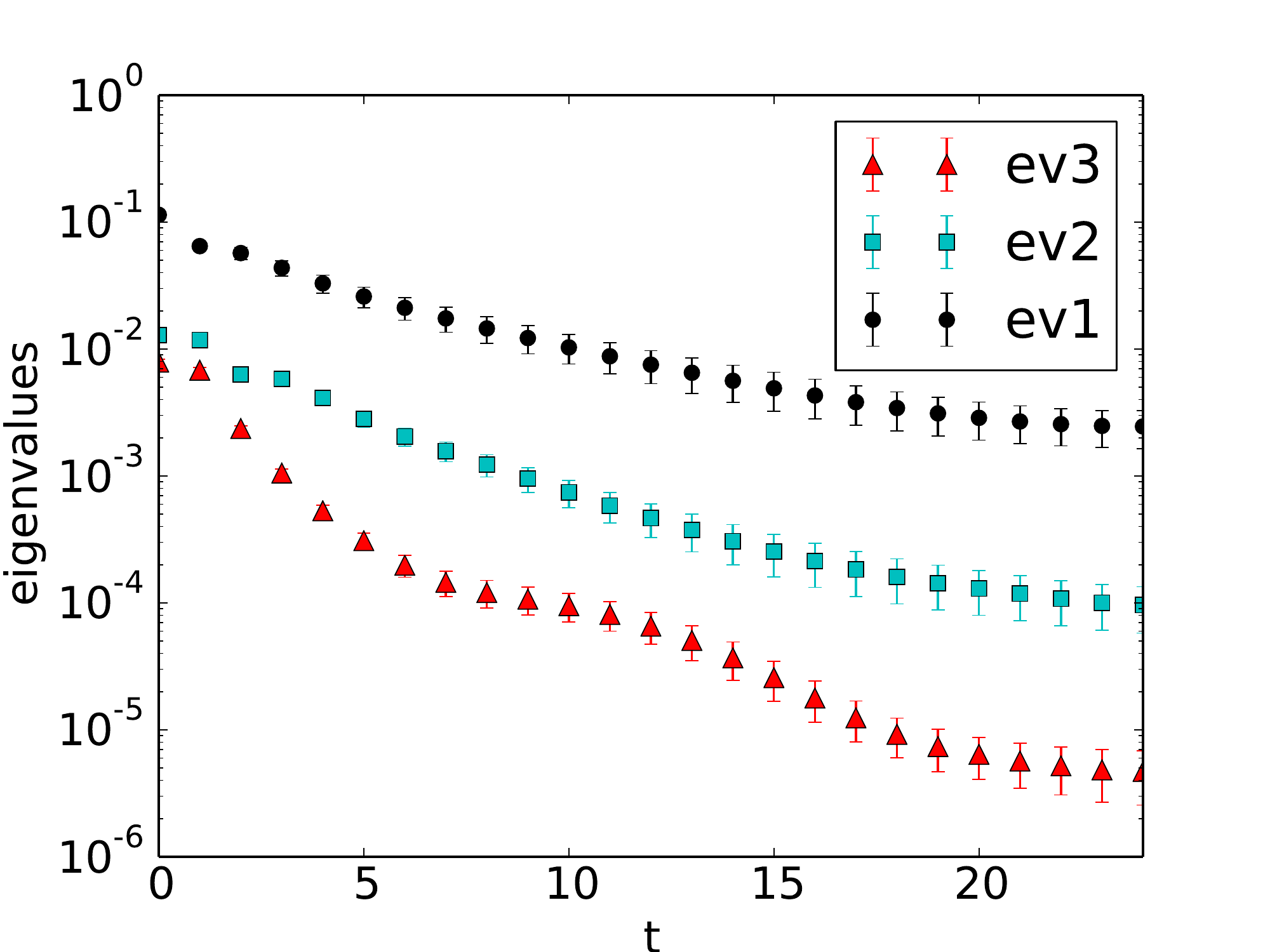}
\small{(b)}\includegraphics[width=8cm,clip]{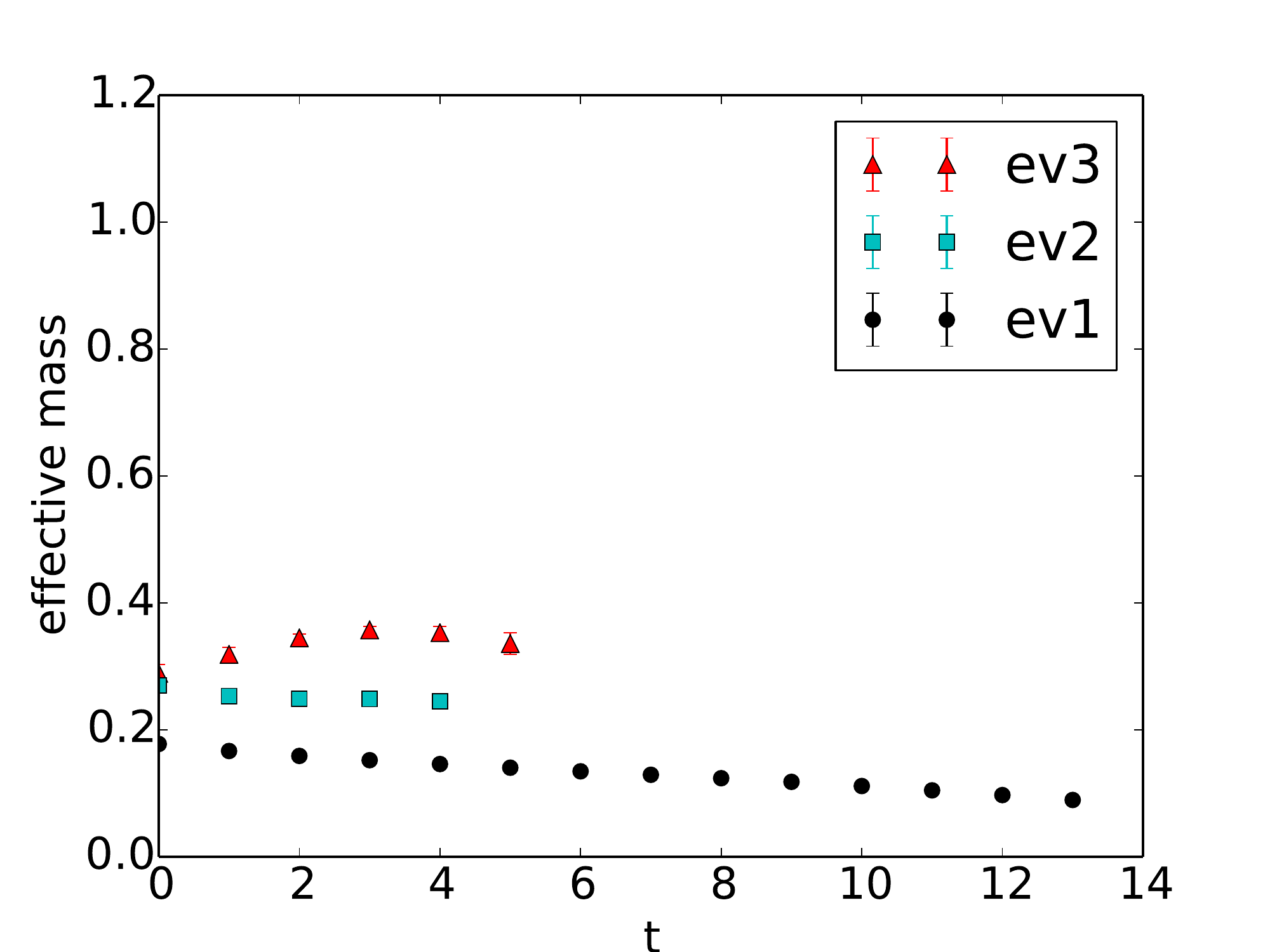}
\caption{ Eigenvalues and effective mass of the correlators for $20 \times 48$ lattice  \label{GN_eigen_meff1}}
\end{center}
\end{figure}

\begin{figure}[htbp]
\begin{center}
\small{(a)}\includegraphics[width=8cm,clip]{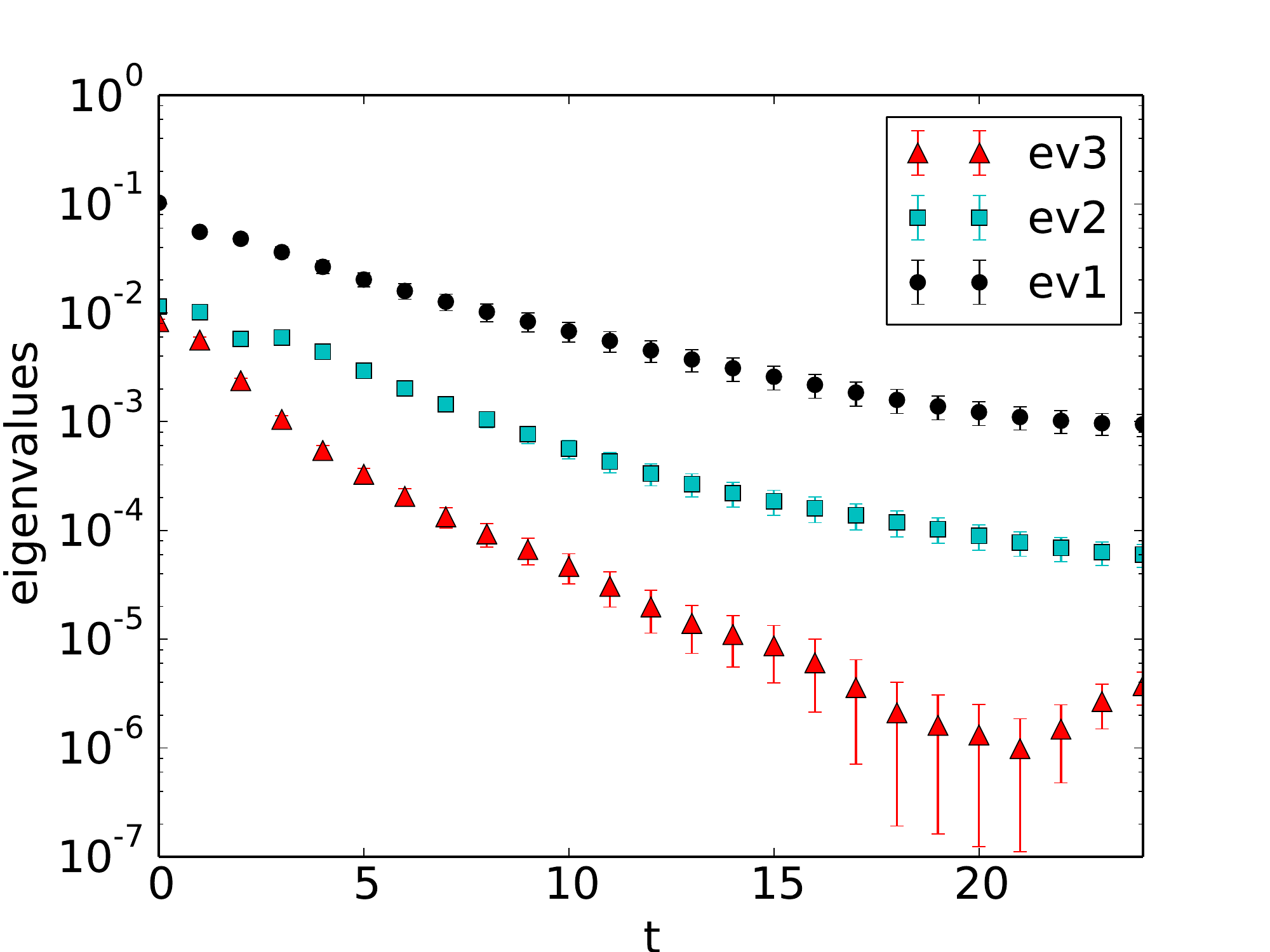}
\small{(b)}\includegraphics[width=8cm, clip]{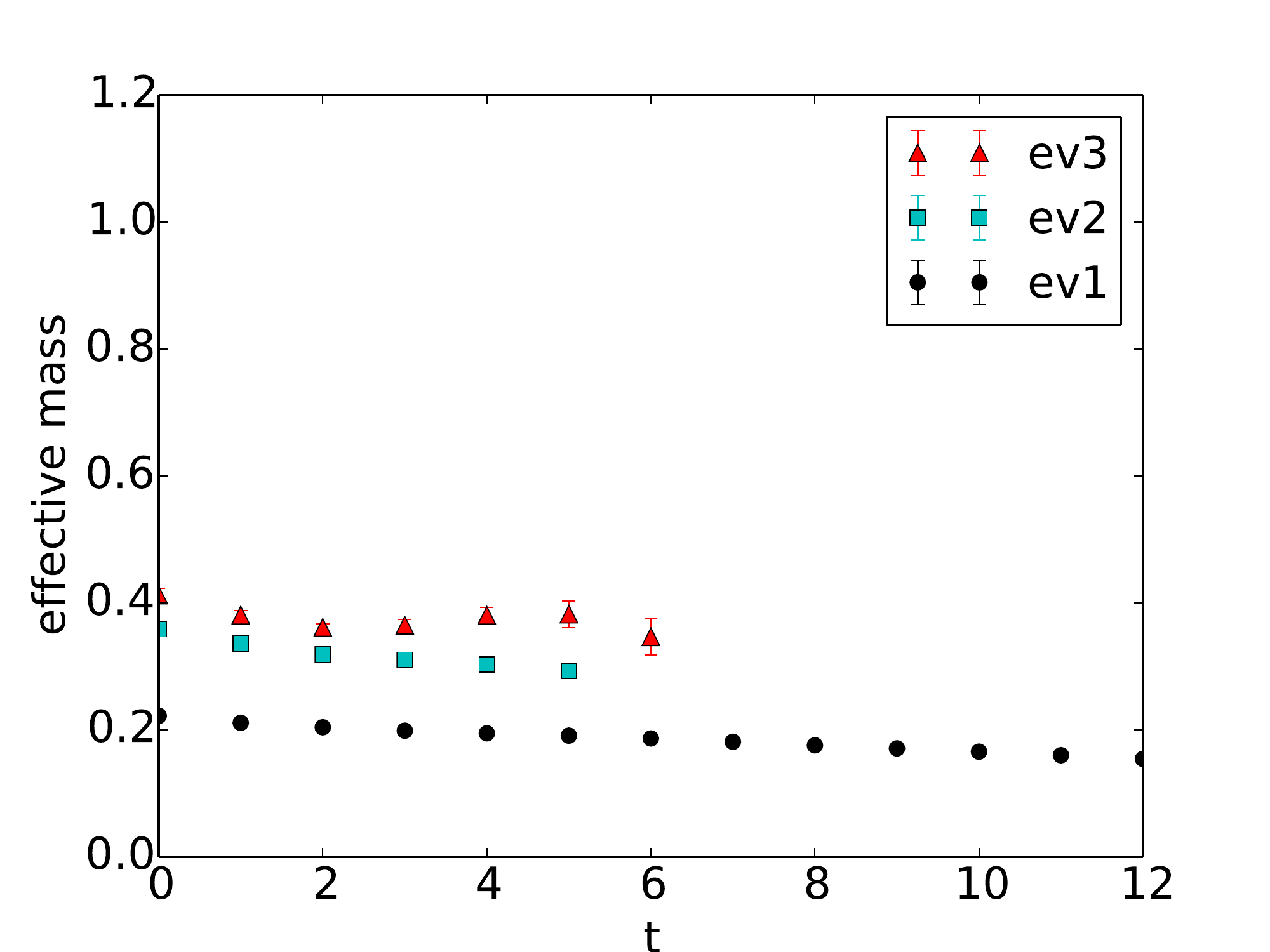}
\caption{ Eigenvalues and effective mass of the correlators for $22 \times 48$ lattice  \label{GN_eigen_meff2}}
\end{center}
\end{figure}

\begin{figure}[htbp]
\begin{center}
\small{(a)}\includegraphics[width=7cm,clip]{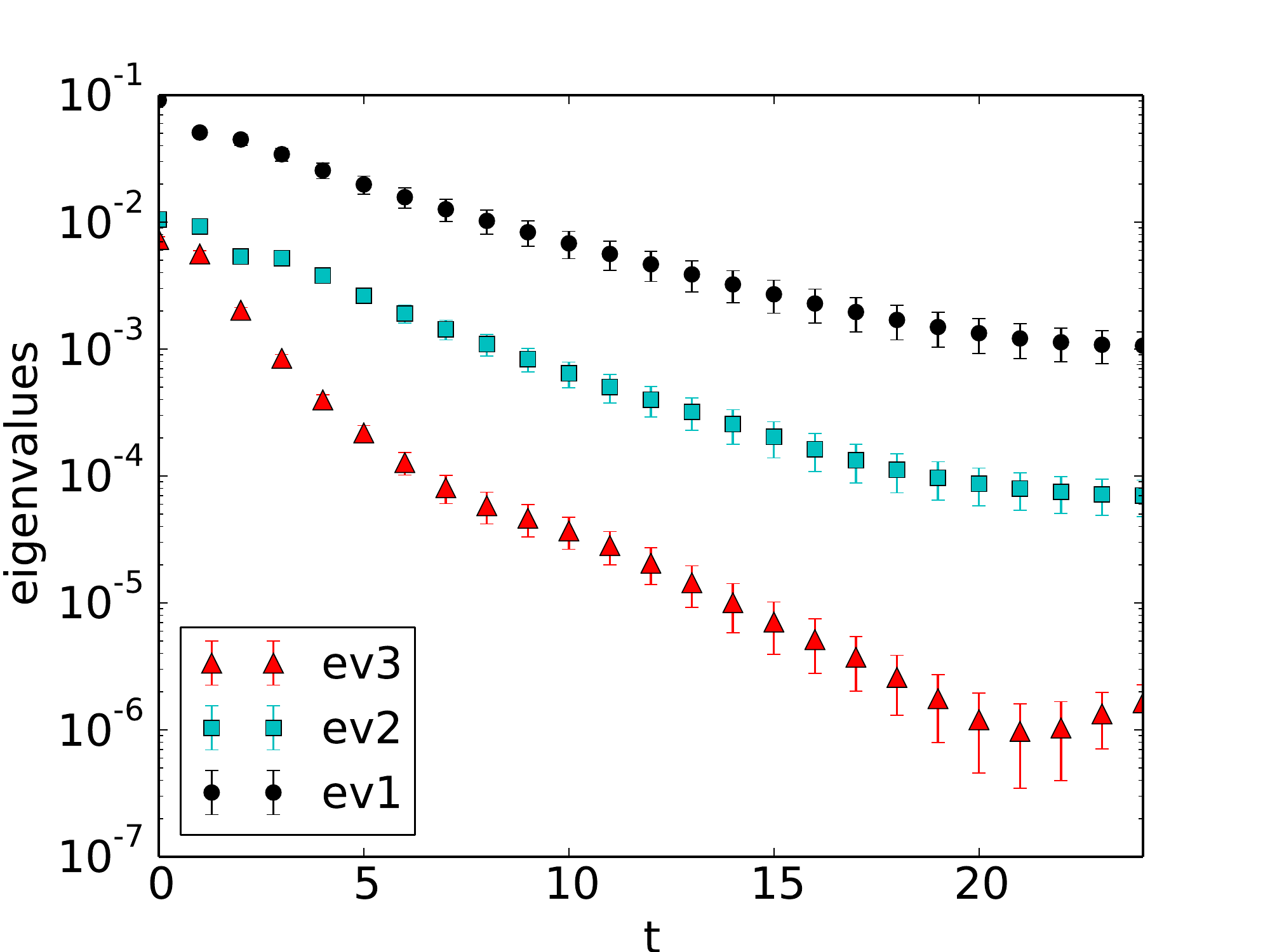}
\small{(b)}\includegraphics[width=7cm, clip]{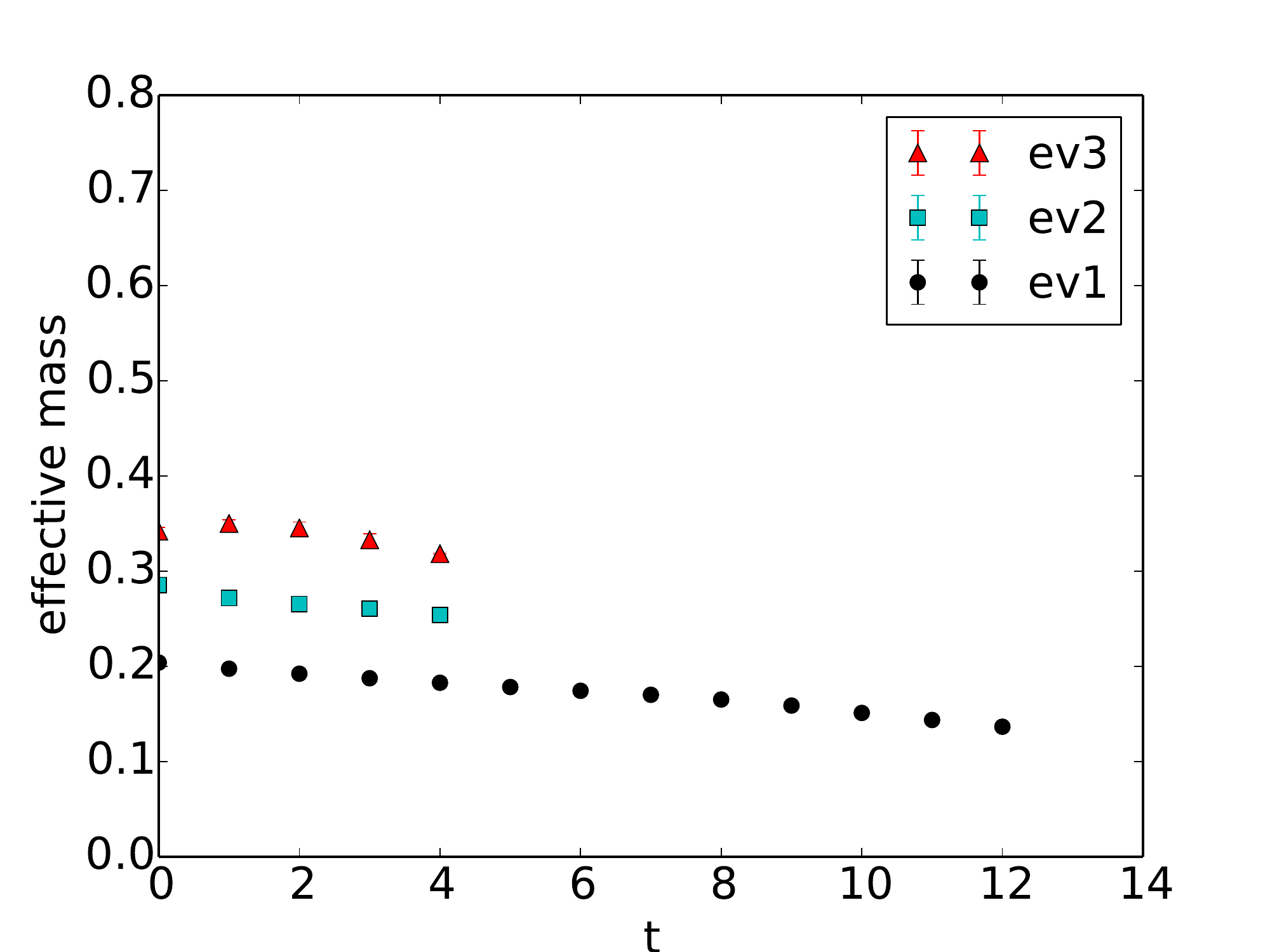}
\caption{ Eigenvalues and effective mass of the correlators for $24 \times 48$ lattice  \label{GN_eigen_meff3}}
\end{center}
\end{figure}

\subsection{Effective mass calculation}
The effective masses are extracted from the correlators at different time slices by  the formula 
\be M_{eff}=\ln \left(\frac{c(t)}{c(t+1)}\right).\label{m_eff}
\ee
 The diagonal correlators $C_{ii}$ and corresponding effective masses are shown in Fig.\ref{GN_meson_corr}(a) and (b) for $m_0=0.03$ and $\beta=0.7$.  The small value of mass is taken to be close to the massless  limit.
Eq.(\ref{m_eff})  is an approximate formula and found good for ground state but  can also produce approximate values for the excited states. 
As can be seen from Fig. \ref{GN_meson_corr}(b), except for the ground state, this procedure cannot  extract the excited states  well enough,  we only see
a  hint of two other possible states.  Analysis of the eigenvalue spectrum of the correlation matrix, on the other hand,  provide a better picture for the meson spectroscopy.
 In the variational method \cite{Michael,Luscher}, 
  to get the mass spectrum from eigenvalues one solves the generalized eigenvalue problem defined by
  \be
  C(t) \vec{v}^{(n)}=\lambda^{(n)}(t)C(t_0)\vec{v}^{(n)}
  \ee
  where $C(t)$ is the $N\times N$ correlation matrix  constructed from $N$ interpolators $O_i,~(i=1,2\cdots, N)$. The $n$-th eigenvalue behaves as
  \be
  \lambda^{(n)}(t)=e^{-(t-t_0)E_n}\Big[1+\mathcal{O}(e^{-(t-t_0)\Delta_n})\Big],
  \ee
  where $E_n$ is the energy of the $n$-th state and $\Delta_n$ is the energy gap between the neighboring states. 
In Fig.\ref{GN_eigen_meff} we show the eigenvalue and effective mass plots by solving the generalized eigen value problem for $16\times 48$ lattice with $O_1,~O_2$ and $O_3$ interpolators.  
We are unable to get any extra stable mass values by increasing the matrix dimension of the correlator basis. So, we work with only $O_1,~O_2$ and $O_3$. 
The results become noisier if the matrix dimension is increased by including more correlators. We have shown the eigenvalues and effective mass plots for
the $16\times 48$ lattice in Fig.\ref{GN_eigen_meff}. Three mass plateau can be observed in Fig.\ref{GN_eigen_meff}(b). 
In Fig.\ref{GN_eigen_meff1} , Fig.\ref{GN_eigen_meff2}  and Fig.\ref{GN_eigen_meff3} we have shown the eigenvalues and effective masses for $20 \times 48,~ 22\times 48$ and $24\times 48$ respectively..
\begin{figure}[htbp]
\centering
\includegraphics[width=8cm,height=7cm,clip]{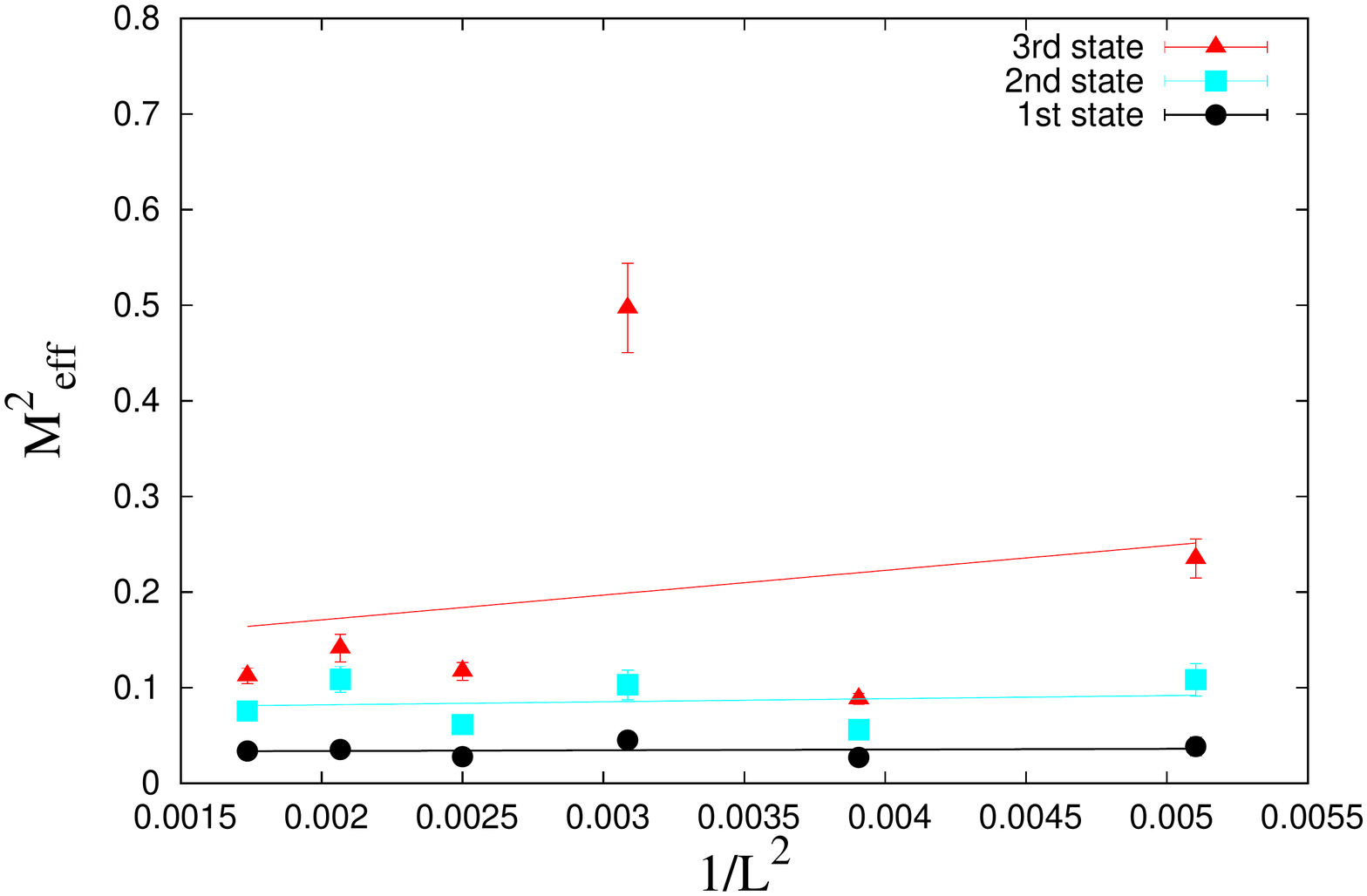}
\caption{\label{volume_dep} Volume dependence of the effective mass.  }
\end{figure}
 In Fig.\ref{volume_dep}, the volume dependence of the effective masses are shown. The ground state and the first excited state show no  volume dependence and hence can  be considered as bound states. The second  excited state however shows a weak  volume dependence. Specially, for $18\times 48$ lattice size, we get an anomalously large mass for the second excited state. The fit for the second excited state shown in Fig.\ref{volume_dep}  includes this anomalous point.  In general, scattering states show strong volume dependence and increase linearly with $1/L^2$, the volume dependency of the second excited state in our case is not very conclusive. 
 But looking at the fit of the points we expect  it to be  a   scattering state.   The results can be contrasted with \cite{Danzer} where the Gross-Neveu model was studied with Wilson fermion. In \cite{Danzer}, except the ground state, all the excited states show strong volume dependence and  are  scattering states. At least for Gross-Neveu model, BC fermion works better for excited state spectroscopy.

\subsection{Fermion mass and chiral phase transition}
\begin{figure}[htbp]
\begin{center}
\small{(a)}\includegraphics[width=8cm,clip]{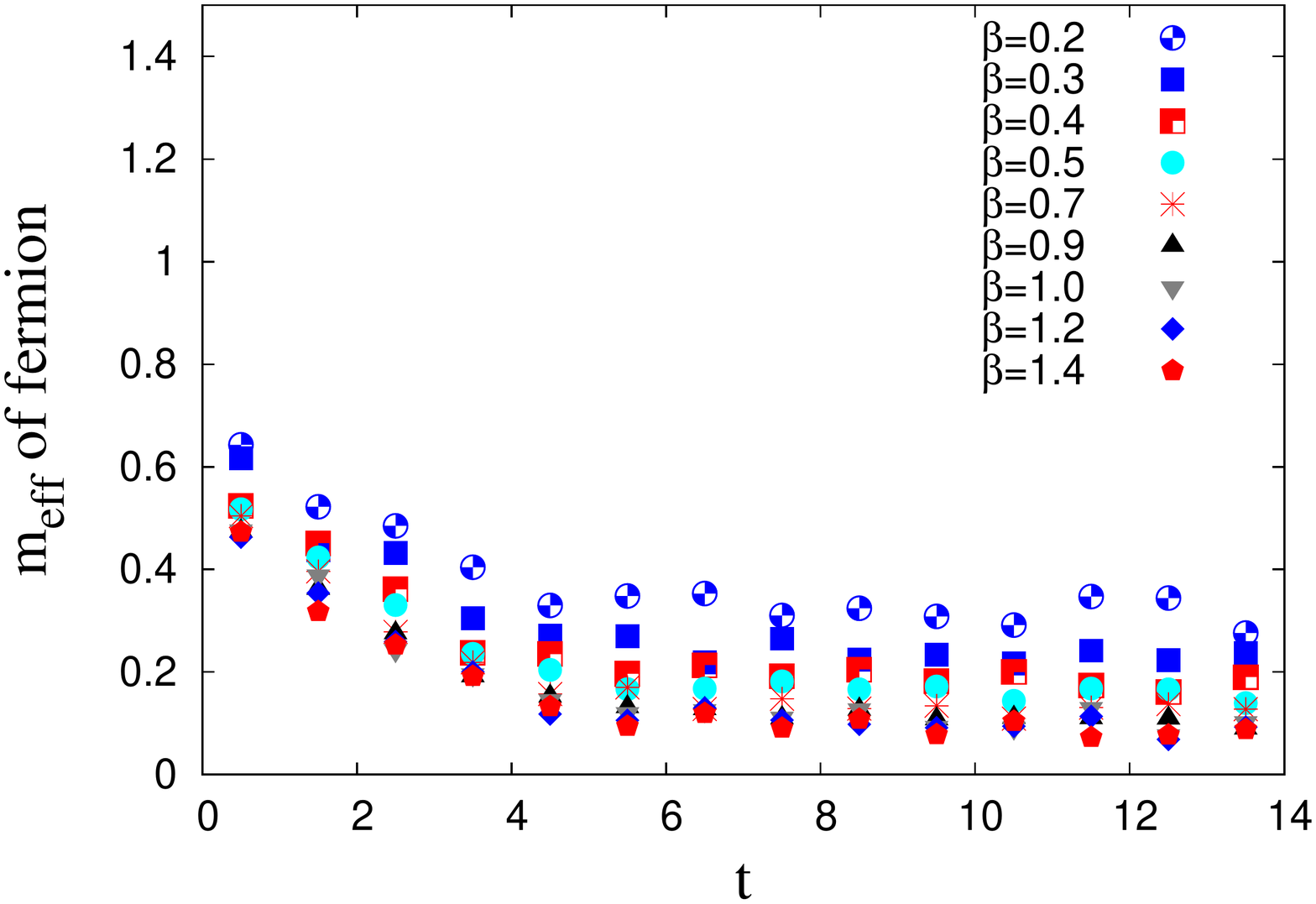}
\small{(b)}\includegraphics[width=8cm,clip]{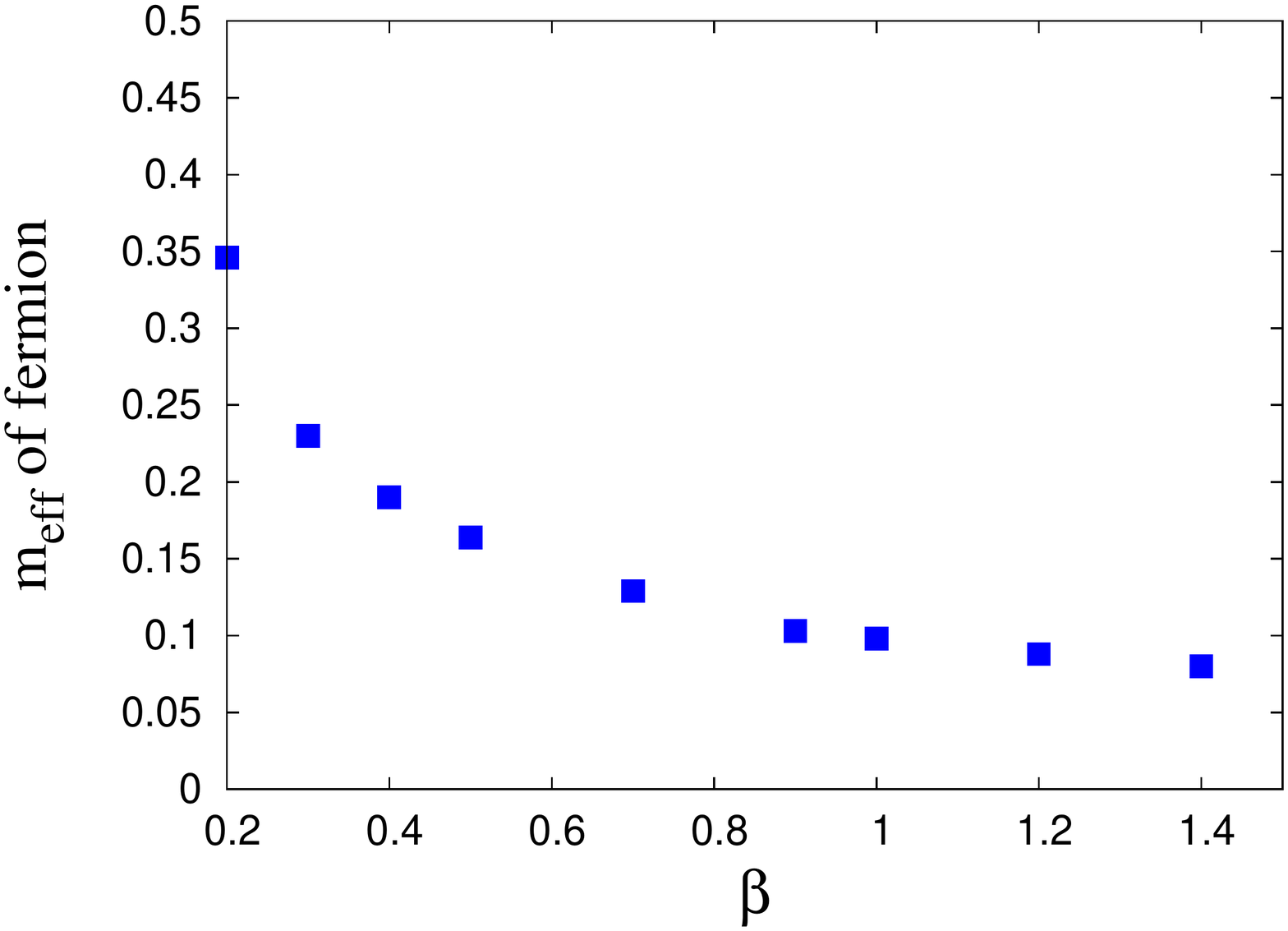}
\caption{\label{elec_mass}(a) Effective mass of the fermion in GN model, (b) the variation of effective electron mass with $\beta$ is consistent  with the chiral phase transition demonstrated in Ref.\cite{GCB}.  The fermion bare mass is $m_0=0.03$.}
\end{center}
\end{figure}

  In GN model, we have also extracted the effective mass of the fermion.  For the fermion mass calculation, we  consider the  correlator with $O(t)=\psi(x,t)$ and 
   evaluate the effective  fermion mass for different  values of the coupling constant $\beta$.  The effective masses for different $\beta$ are   shown in Fig.\ref{elec_mass}(a).  
   We take $c_3=0.0001$ to be close to the phase boundary.
  Fig. \ref{elec_mass}(b) shows the variation of effective electron mass with the coupling constant. At small $\beta$( i.e., at large coupling), the electron mass rapidly increases   indicating a phase transition as can be seen from Fig. \ref{elec_mass}(b). In \cite{GCB}, it was shown with Borici-Creutz fermion that the GN model with a discrete chiral symmetry shows a second order chiral phase transition at $\beta \approx 0.4$. The current result for the $\beta$-dependence  of the electron mass   is consistent with that finding. The systematics of the continuum limit are not studied here.
 
\section{Meson in 2D QED}
\begin{figure}[htbp]
\centering
\small{(a)}\includegraphics[width=8cm,clip]{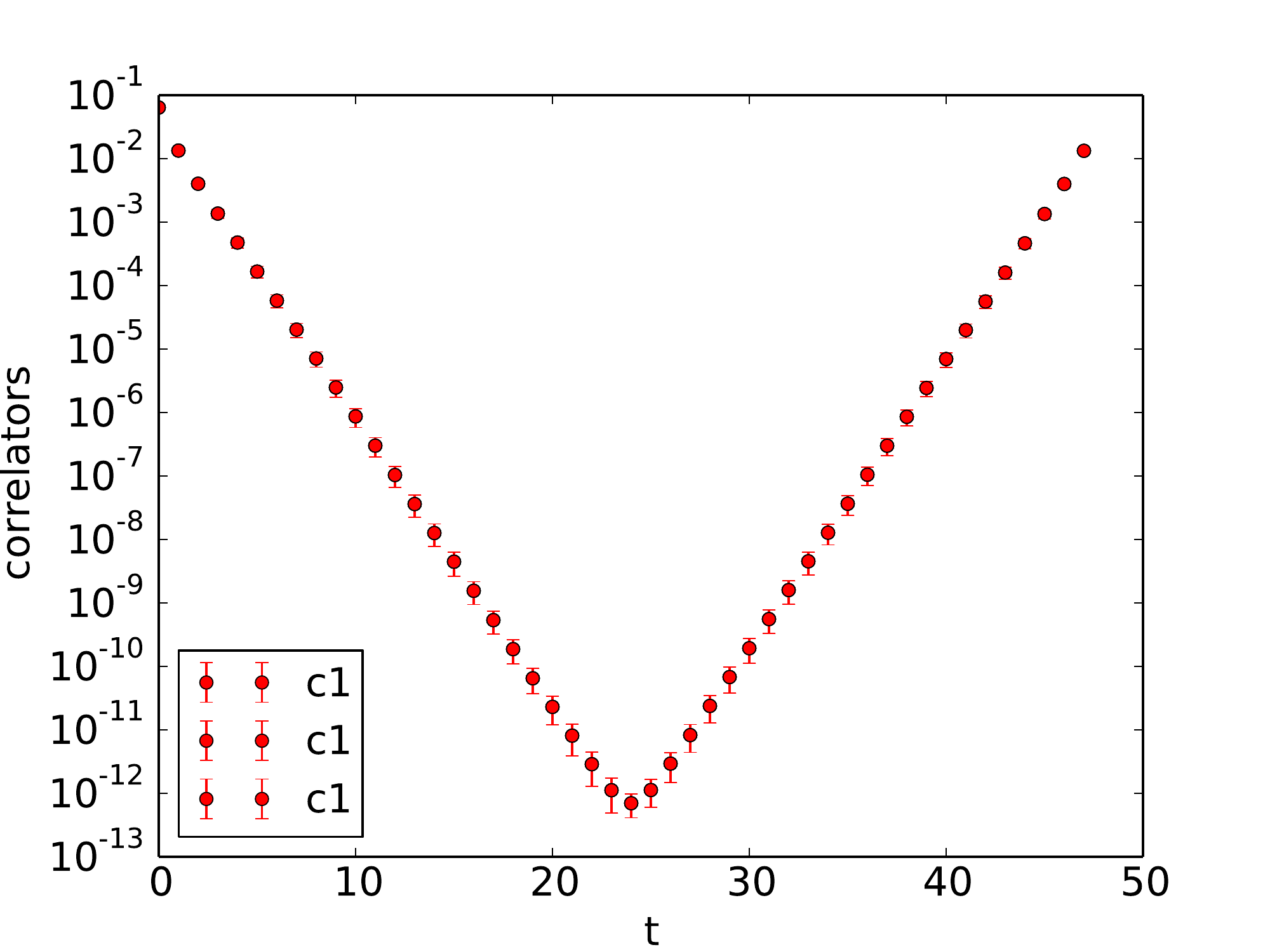}
\small{(b)}\includegraphics[width=8cm,clip]{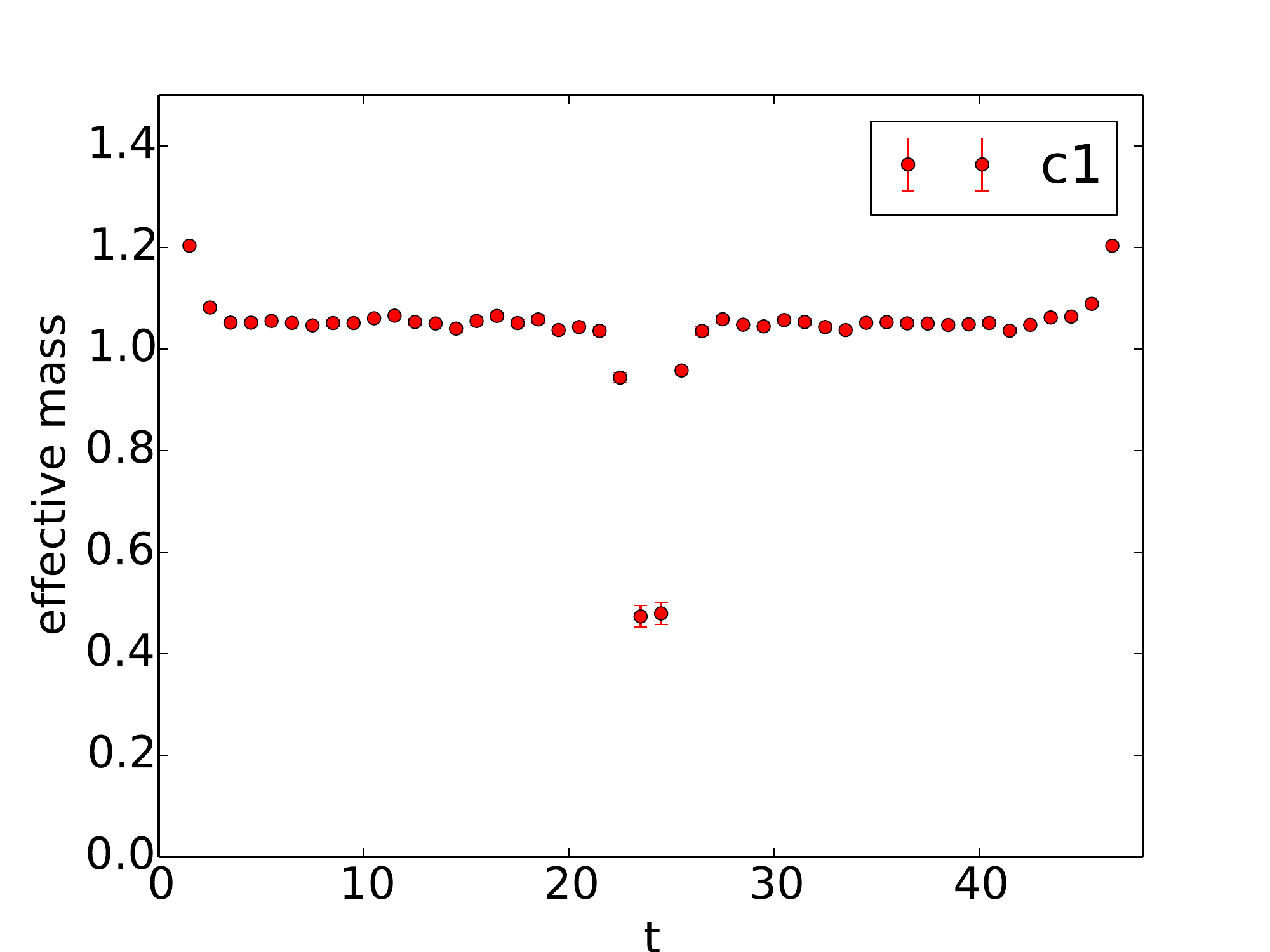}
\caption{Effective mass of meson in 2D QED for $m_0=0.05$ and $\beta=0.3$.\label{QED_meson}}
\end{figure}

\begin{figure}[htbp]
\centering
\includegraphics[width=8cm,clip]{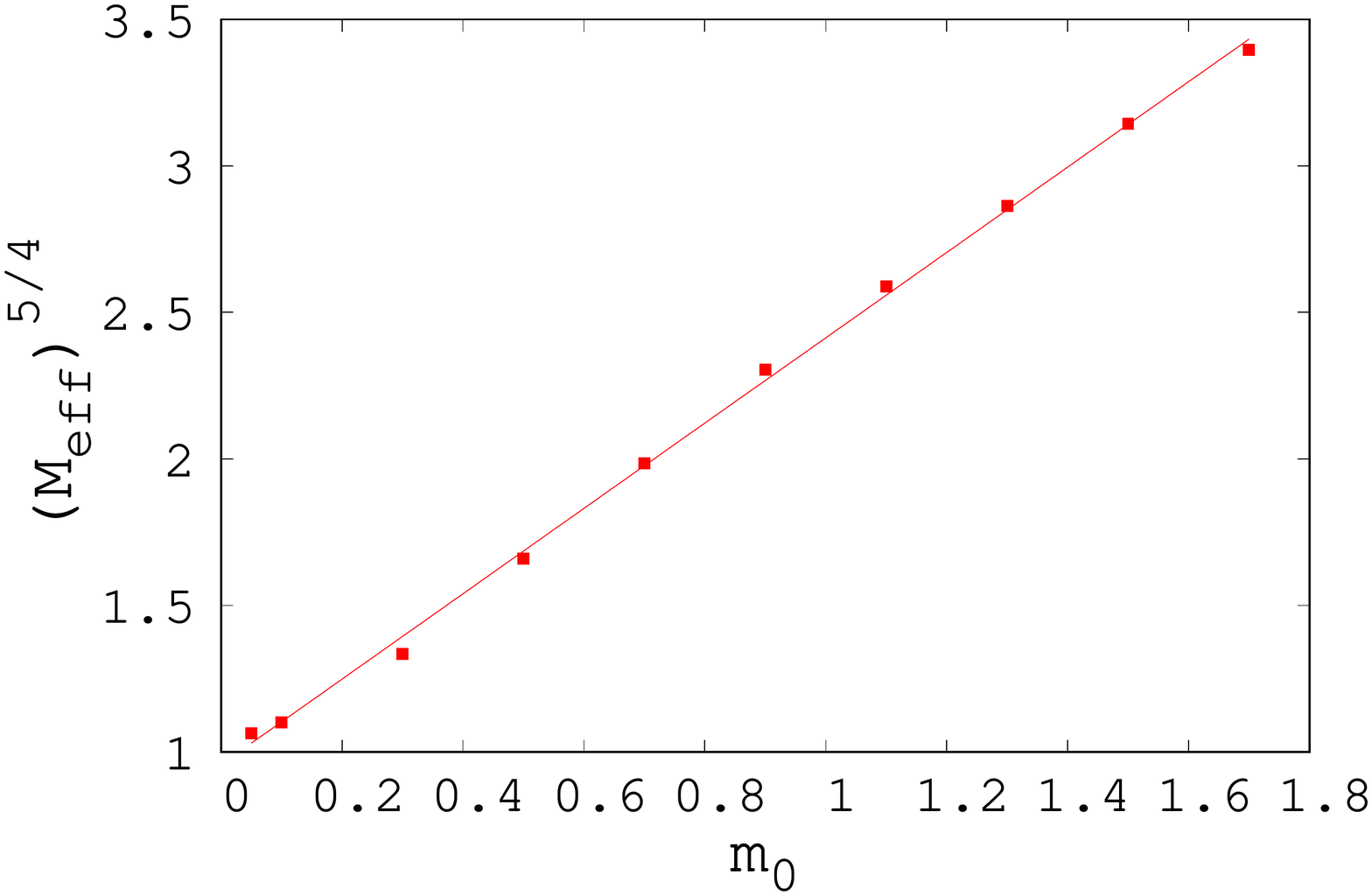}
\caption{ Fermion mass dependence of the effective pion mass in QED$_2$ for a fixed $\beta=0.3$. 
\label{QED_mass_dep}}
\end{figure}

In this section, we extend the study of spectroscopy with BC fermion formulation to gauge theory.  For this purpose, we 
implement the BC fermion in  a 2D $U(1)$ gauge theory  and  extract the meson masses.  QED in 2D is also a confined theory and serves as a good toy model for QCD. The Gross-Neveu model, having a discrete chiral symmetry, undergoes spontaneous breaking but in the massless Schwinger model the chiral symmetry is continuous and cannot be spontaneously broken.
The lattice action with BC fermion reads, 
\be
S=\beta\sum_p[1-\frac{1}{2}(U_p+U^{\dagger}_p)]+\phi^{\dagger}(D^{\dagger}D)^{-1}\phi.
\ee  where $U_p$ is the Wilson Plaquette action with
 \be 
 U_p=U_{i,\mu}U_{i+\mu,\nu}U^{\dagger}_{i+\nu,\nu}U^{\dagger}_{i,\nu}.
 \ee where, $i$ is the site index and $\mu,\nu$ are the directions
and $D$ is the BC Dirac operator defined in eqn.(\ref{bcdirac}).
 After  including gauge fields we get,
 \be 
  D_{mn}&=&\frac{1}{2}(\gamma_\mu+i(\Gamma-\gamma_\mu))U_{\mu}(n-\mu)\delta_{n,m+\mu}-\nonumber\\
  &&\frac{1}{2}(\gamma_\mu-i(\Gamma-\gamma_\mu))U^{\dagger}_{\mu}(n)\delta_{n,m-\mu}-((2-c_3)i\Gamma-m_0)\delta_{m,n}.
 \ee
Here we concentrate only for the lowest lying meson mass. The correlator with operator $O_1(t)$ couples to the ground state and provides the mass for the lowest  state  which we call pion following the general convention.
 In Fig.\ref{QED_meson}(a) we have shown the correlator at different time slices and the effective meson mass in 2D QED.  The results are presented for the fermion (electron) mass $m_0=0.05$ and  $\beta=0.3$.  The effective  pion mass is shown in Fig.\ref{QED_meson}(b).
 The  Schwinger model in continuum  can be written as a  bosonic theory.   The pion mass in the bosonized theory can be exactly calculated\cite{smilga} and   for $m_0\ll g$, can be written as
 \be
 M_{eff}&=&A~ m_0^{N_f/(N_f+1)} g^{1/(N_f+1)} = A~ m_0^{5/4} g^{1/5}~~ \rm{(for~ N_f=4)},\\
 &=& A~ m_0^{4/5} \beta^{-1/10}\nonumber
 \ee
 where $A$ is a constant and $\beta=1/g^2$.   In Fig.\ref{QED_mass_dep}, we show the fermion mass dependence of the effective pion mass ($M_{eff}^{5/4}\propto m_0$). The plot is done for  small $\beta$ so that $m_0$ is always less than $ g$.  The lattice data are in
  well agreement with the analytic result.

\section{Summary}
Minimally doubled fermions may provide an  efficient lattice formalism to study chiral fermion which  is expected to be computationally cheaper than the other existing lattice formalisms. Since, both the minimally doubled fermion formulations (KW and BC) break  hypercubic symmetries on the lattice, they require non-covariant counter terms. Only detailed numerical studies can confirm how bad or manageable  its effects are on the lattice,  and whether any meaningful computation is possible with minimally doubled fermion or not.  In this work, we have studied  the BC fermion  in some simple models.
 We have extracted the excited state  mass  spectrum in  Gross-Neveu model using BC fermion. The absence of volume dependence of  the first two states suggests that they are   mesonic bound states and the third state show a weak volume dependence. The effective fermion mass has also been evaluated as a function of the coupling constant $\beta$. It shows a phase transition consistent with the result obtained in \cite{GCB}. We have also evaluated the lowest lying  meson mass in QED$_2$.  The lattice results are consistent with the prediction from analytic calculation in the bosonized model.
  Our investigations suggest that  BC fermion formalism might be a promising  alternative to study the chiral fermions on a lattice. One obviously needs more detailed numerical study in 4D gauge theory with  dynamical BC fermion to confirm that claim. Invesigation of BC fermion in 4D QCD  in a mixed-action lattice simulation is in progress.


\end{document}